\documentclass[12pt]{article}
\usepackage{graphicx}
\newcommand{\dbox}{\,\raise2pt\hbox{\fbox{\rule{2.5pt}{0pt}\rule{0pt}{2.5pt}}}\,}
\newcommand{\qed}{\,\raise0pt\hbox{\mbox{\rule{6.5pt}{6.5pt}}}}
\newcommand{\simlt}{\raise3pt\hbox{\,\scriptsize$<$}\kern-6.5pt\lower2pt\hbox{\scriptsize$\sim$}\,}

\newcommand{\ket}[1]{\mbox{$| #1 \rangle$}}

\begin{document}
\setlength{\baselineskip}{7mm}

\begin{titlepage}
 \begin{normalsize}
  \begin{flushright}
        UT-Komaba/06-13\\
        November 2006
  \end{flushright}
 \end{normalsize}
 \begin{LARGE}
   \vspace{1cm}
   \begin{center}
       Level Truncated Tachyon Potential in Various Gauges\\
   \end{center}
 \end{LARGE}
  \vspace{5mm}
 \begin{center}
    Masako {\sc Asano}
            \hspace{3mm}and\hspace{3mm}
    Mitsuhiro {\sc Kato}$^{\dagger}$
\\
      \vspace{4mm}
        {\sl Faculty of Liberal Arts and Sciences}\\
        {\sl Osaka Prefecture University}\\
        {\sl Sakai, Osaka 599-8531, Japan}\\
      \vspace{4mm}
        ${}^{\dagger}${\sl Institute of Physics} \\
        {\sl University of Tokyo, Komaba}\\
        {\sl Meguro-ku, Tokyo 153-8902, Japan}\\
      \vspace{1cm}

  ABSTRACT\par
 \end{center}
 \begin{quote}
  \begin{normalsize}
New gauge fixing condition with single gauge parameter proposed by the authors is applied to the level truncated analysis of tachyon condensation in cubic open string field theory. It is found that the only one real non-trivial extremum persists to appear in the well-defined region of the gauge parameter, while the other solutions are turned out to be gauge-artifacts. Contrary to the previously known pathology in the Feynman-Siegel gauge, tachyon potential is remarkably smooth enough around Landau-type gauge.
\end{normalsize}
 \end{quote}

\end{titlepage}
\vfil\eject
\section{Introduction}

Stimulated by the Sen's conjecture~\cite{Sen:1999xm}, string field theoretical study of tachyon condensation has been carried out in depth in these several years. In particular, the level truncation approximation was powerful enough to show numerically the potential depth of the vacuum being equal in good accuracy to the D25-brane tension in bosonic open string field theory~\cite{Kostelecky:1989nt,Sen:1999nx,Moeller:2000xv}.

Almost all of these previous studies are conducted in Siegel gauge which has been practically unique choice in string field theory for these twenty years. This gauge, however, is known to show a pathological behavior in effective tachyon potential. Namely, branch points appear in both side of larger and smaller field values of the tachyon so that one could not go beyond this small region. Fortunately previous study showed that perturbative vacuum and candidate true vacuum are both inside this region at available order of level truncation, though the higher an approximation went up the smaller its smooth region became.
A preceding study~\cite{Ellwood:2001ne} suggested that these behavior came from the gauge boundary where the gauge slice became not to cross the gauge orbit. In order to confirm these observations and to avoid the problem new covariant gauge other than Siegel gauge has been desired for a long time.

The present authors recently proposed~\cite{AK} a new single-parameter family of covariant gauges in string field theory which includes Siegel gauge at a special point. This gauge family is a natural extension of the covariant gauges in ordinary gauge theory to the string field theory so that the gauge parameters of them have a natural correspondence with each other.

The purpose of the present paper is to apply this new gauge to the tachyon condensation in level truncation approximation. We will investigate the gauge (in-)dependence of the previously known behavior of tachyon condensation and show manifestly which is physical or gauge artifacts. Then it turns out that there are large parameter region where the potential behaves smooth enough. Also in at least level 2 approximation, there are only two extrema which persist to exist independently of the gauge; one is the perturbative vacuum and the other is the candidate tachyon vacuum.

It turns out that the Siegel gauge is sitting in a subtle area where the influence from the gauge horizon is still too large to obtain the whole shape of the potential, while the energy of the vacuum does not so much deviate from the gauge independent value.

In the following, after setting up the necessary ingredients for the level truncation analysis including our gauge fixing conditions, we investigate the effective tachyon potential in Section~2. In particular, gauge dependence of the potential in level (2,6) truncation will be throughly described and the smooth behavior in the Landau-type gauge will be shown. In Section~3 the gauge dependence of the vacuum energy is analyzed. Section~4 is devoted to the discussions.

\section{Setup: the action and gauge fixing conditions}
In this section we briefly describe our basic setup for the calculation of the tachyon potential in level truncation\footnote{See review article, e.g. Ref.\cite{Taylor:2003gn}, for the fundamentals.} and the gauge fixing conditions proposed in Ref.\cite{AK}.

The action for the cubic open string field theory~\cite{Witten:1985cc}
is given by
\begin{equation}
S = -{1 \over 2} \langle \Phi_1, Q \Phi_1  \rangle
-{g \over 3} \langle \Phi_1, \Phi_1 \ast \Phi_1  \rangle 
\label{eq:S}
\end{equation}
where $\Phi_1$ is the string field of ghost number $N^g=1$
and $g$ is the coupling constant.
This action is invariant under the gauge transformations
\begin{equation} 
\delta \Phi_1 = Q \Lambda_0 + 
g (\Phi_1 \ast \Lambda_0 - \Lambda_0 \ast \Phi_1)
\label{eq:gaugetr}
\end{equation}
where the BRST operator $Q$ is given by
\begin{equation}
Q = \tilde{Q} + c_0 L_0 + b_0 M 
\end{equation}
with 
\begin{equation}
\tilde{Q}= \sum_{n\ne 0} c_{-n} L_n^{({\rm m})}  
-\frac{1}{2} \sum_{\stackrel{mn\ne 0}{\mbox{\tiny $ m\!+\!n \ne 0$}}} (m-n)\,:c_{-m} c_{-n} b_{n+m}: ,
\qquad 
M=-2 \sum_{n>0} n c_{-n} c_{n} .
\end{equation}

To define the gauge fixing conditions of Ref.\cite{AK}, we write the string field as $\Phi_1=\phi^{(0)}+c_0 \omega^{(-1)}$ such that $\phi^{(0)}$ and $\omega^{(-1)}$ are expanded by the states without $c_0$. Here the superscripts $(0)$ or $(-1)$ denote the ghost number of non-zero modes: $\tilde{N^g}=\sum_{n>0}(c_{-n}b_n-b_{-n}c_n)$.
As was shown in Ref.\cite{AK}, the gauge invariance of the action $S$ for $g=0$ is consistently 
fixed by the condition 
\begin{equation}
(b_0 M + a b_0c_0 \tilde{Q}) \Phi_1 =0 
\quad(\Leftrightarrow \quad M \omega^{(-1)} + a\tilde{Q} \phi^{(0)}=0) 
\label{eq:gca}
\end{equation}
for a real parameter $a\ne 1$ including $a = \pm\infty$. 
In the following, we take all of these gauge conditions including the $a=1$ case to analyze the action $S$ since we have no a priori reason to exclude $a=1$ case unless $g=0$. 
Note that the break-down of each gauge fixing condition would occur at a certain different region of the configuration space in general.
As described in Ref.\cite{AK}, $a=0$ is equivalent to the Feynman-Siegel gauge and $a=\infty$ corresponds to the Landau gauge for the massless gauge mode.

The gauge fixed action for each of the above gauge conditions contains string fields of all the ghost number so as to incorporate infinite sequence of ghost and anti-ghost fields introduced in the gauge fixing procedure~\cite{AK}. 
We will, however, put all string fields with $N^g\ne 1$ (i.e., all ghost and anti-ghost fields) to zero in the following since our analysis is limited to the classical one;
We just impose each gauge condition (\ref{eq:gca}) on the action (\ref{eq:S}) and use it as gauge fixed action for the gauge condition.

In the following analysis, we will concentrate on the analysis of the tachyon condensation problem and restrict the states in the string field to zero-momentum scalar fields with even level. This restriction is justified by twist symmetry of the action and the fact that we only consider Lorentz invariant configurations%
\footnote{We do not further restrict to the so called universal subspace~\cite{Sen:1999nx,Rastelli:2000iu} whose matter part is limited to the states $\ket{0}$ and its descendants of matter Virasoro algebra $L_{-n}^{\rm (m)}$.
}.
Under the restriction, the string field is expanded as 
\begin{equation}
\Phi_1 = \phi \ket{\!\downarrow} + \sum_{i} \phi_i \ket{f_i} + c_0 \sum_{j} \omega_j \ket{g_j} 
\label{eq:Philt}
\end{equation}
where $\phi$ is the zero-mode tachyon field with $\ket{\!\downarrow} = c_1\ket{0}$, 
$\ket{f_i}$ and $c_0 \ket{g_i} $ are ghost number $1$ scalar states of even level, and $\phi_i$ and $\omega_j$ are corresponding scalar fields.
Note that our gauge fixing condition can be applied consistently on the restricted subspace of the state space 
since the condition is covariant and is closed within the states of the same level.
Also, there are relations
$ \{ \tilde{Q} \ket{f_i}\} = \{M\ket{g_j}\}$
and the isomorphism $ \{ \ket{g_i} \} \sim  M \{\ket{g_i} \}$ 
since $L_0 = L -1 \ne 0$ for each state in the subspace with level $L$.
For the general background of the relations, see Ref.\cite{AK}.  

Let $\Phi_1^{(L)}$ denotes the string field (\ref{eq:Philt}) expanded by the states up to level $L$.
The effective potential up to level $L$ is given by substituting $\Phi_1^{(L)}$ into the action (\ref{eq:S}) as
\begin{equation}
V^{(L,3L)}(\phi, \{\phi_i\},\{\omega_j\}) = 
-S \left(\Phi_1^{(L)} \right) .
\end{equation}
The gauge fixed form of the effective potential $V_a$ is derived from the $V^{(L,3L)}$ by imposing the gauge fixing condition up to level $L$, which is obtained by substituting $\Phi_1=\Phi_1^{(L)}$ into (\ref{eq:gca}).
For $|a|<\infty$, all the $\omega_j$ fields are written by the linear combinations of $\phi_i$'s. For $a=\infty$, only $\phi_i$ fields are restricted by the condition $\tilde{Q}\phi^{(0)}=0$ so that we make $\{ \phi'_{i'} \}$ denote  
the set of states $\phi_i$ satisfying the condition. 
Thus, the effective potential for each gauge $|a|<\infty$ 
(or $|a|=\infty$) is written as a function of $(\phi,\phi_i)$ 
(or $(\phi,\phi'_{i'},\omega_j)$) as
\begin{equation}
V^{(L,3L)}_{a}= V^{(L,3L)}_{a} (\phi,\phi_i) ,\qquad
V^{(L,3L)}_{\infty} = V^{(L,3L)}_{\infty} (\phi,\phi'_{i'},\omega_j).
\end{equation}
In the following, we sometimes represent $\psi_i=\phi_i$ for $|a|<\infty$ and $\psi_i=(\phi'_{i'}, \omega_j)$ for $|a|=\infty$ and 
write simply $V^{(L,3L)}_{a}=V^{(L,3L)}_{a}(\phi,\psi_i)$ for every $|a|\le \infty$.

Let us take an example of $L=2$. The string field up to this level 
is given as 
\begin{equation}
\Phi_1^{(L=2)} =  \phi \ket{\!\downarrow} 
+ \phi_1 (\alpha_{-1} \cdot \alpha_{-1}) \ket{\!\downarrow} 
+ \phi_2 b_{-1} c_{-1} \ket{\!\downarrow} 
+ \omega_1 c_0 b_{-1} \ket{\!\downarrow}. 
\label{eq:sfL2}
\end{equation}
By substituting this to $-S$, we obtain the effective potential $V^{(2,6)}(\phi,\phi_1,\phi_2,\omega_1)$ in the level (2,6) truncation. 
The approximate gauge transformation up to this level is given by the use of $\Phi_1=\Phi_1^{(L=2)}$ and $\Lambda= \lambda b_{-2} \ket{\!\downarrow}$ in (\ref{eq:gaugetr}). 
The gauge fixing condition (\ref{eq:gca}) up to this level is represented by an equation
\begin{equation}
- 4 \omega_1 + a (26 \phi_1 + 3 \phi_2 )  = 0 .
\end{equation}
The effective potential $V_a$ for each gauge condition is given by imposing the condition on $V^{(2,6)}(\phi,\phi_1,\phi_2,\omega_1)$ as
\begin{eqnarray}
V^{(2,6)}_{a}(\phi,\phi_1,\phi_2) &=& 
 V^{(2,6)}(\phi,\phi_1,\phi_2,\omega_1\!=\! a(26 \phi_1 + 3 \phi_2 )/4  ) ,
\\
V^{(2,6)}_{\infty}(\phi,\phi_1,\omega_1) &=& 
 V^{(2,6)}(\phi,\phi_1,\phi_2\!=\! -{26}\phi_1/3 ,\omega_1 ).
\end{eqnarray}

The effective tachyon potential in level $(L,3L)$ truncation for each gauge $a$  
is obtained by solving equations of motion of $V_a(\phi, \psi_i)$ 
with respect to all the fields except $\phi$ as
\begin{equation}
V^{(L,3L)}_a(\phi) = V^{(L,3L)}_a (\phi, \psi_i\!=\!\psi^{\rm sol}_i(\phi) ).
\end{equation}
Here $\psi^{\rm sol}_i(\phi)$ is a solution of 
$\{ \frac{\partial}{\partial \psi_i} V_a(\phi,\psi_i) =0 \}$ for a fixed $\phi$.
For the level $(0,0)$ truncation where the only $\phi$ field is concerned, the effective tachyon potential is given by $V^{(0,0)}(\phi)$ itself:
\begin{equation}
V^{(0,0)}(\phi) = - \frac{1}{2}\phi^2 + g \bar{\kappa} \phi^3,
\label{eq:V00}
\end{equation}
where 
$$\bar{\kappa}=\frac{1}{3}\left(\frac{3\sqrt{3}}{4} \right)^3.$$
In general, there exist a number of real solutions $\psi^{\rm sol}_i(\phi)$ at higher levels and correspondingly there appear a number of branches for $V_a^{(L,3L)}(\phi)$.

We will analyze the properties of $V_a^{(L,3L)}(\phi)$ in the next section.
By the Sen's conjecture~\cite{Sen:1999xm}, the tachyon effective potential $V(\phi)$ should have a non-trivial extremum where the potential value coincides with minus of the tension of D25-brane (times volume of space-time) 
\begin{equation}
-T_{25}=- \frac{1}{2\pi^2g^2}.
\end{equation}
We also analyze the gauge dependent properties of the vacuum solution in the level truncation by solving the equations of motion of $V_a(\phi, \psi_i)$.

\section{Effective tachyon potential}

\subsection{Detailed analysis in level (2,6)}

In the level 2 we have four fields before gauge fixing and the gauge condition removes one field, so only three fields $(\phi, \psi_1, \psi_2)$ are relevant. This makes computation much easier so that we can study this level to a great extent.

\begin{figure}[htbp]
\begin{center}
\begin{picture}(450,600)(0,0)
\put(-80,-150){\includegraphics{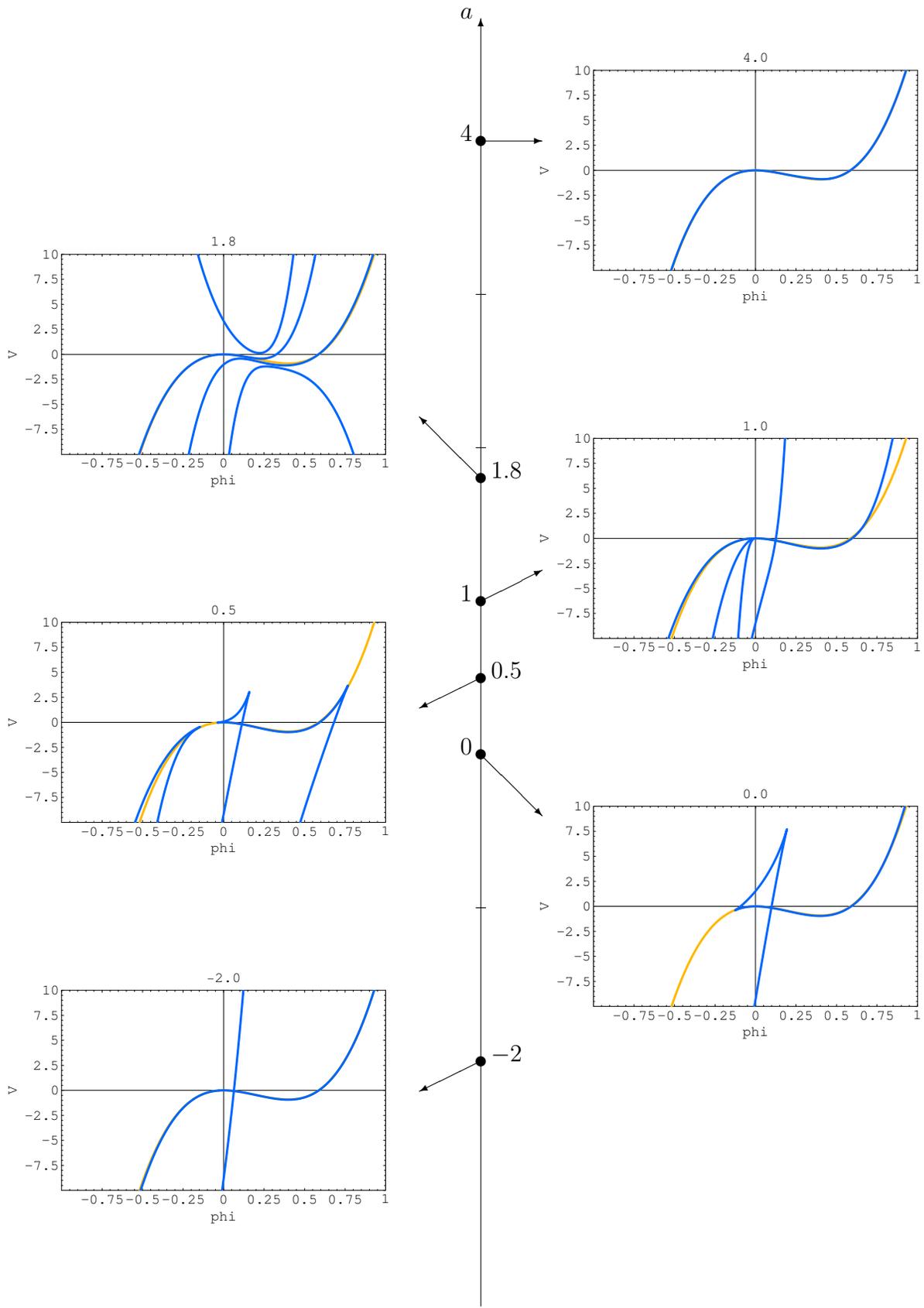}}
\end{picture}
\caption{{\bf Effective tachyon potential at various $a$}}
\label{fig1}
\end{center}
\end{figure}

First we show the effective tachyon potential at various values of $a$ depicted in Fig.\ref{fig1}. In each plot we superimpose the potential (blue curve) over the physical branch of $a=\infty$ (orange curve) which is helpful as a guide to eye in order to recognize a common tendency of each physical branch.
Note that throughout all figures we use normalized variables $g\bar{\kappa}\phi$ as ``{\tt phi}'' and $V/T_{25}$ as ``{\tt V}''.

At $a=4$ the potential completely overlaps with the orange curve and this branch is stable for larger $a$. The same behavior can be seen in $a=-2$ in which another independent branch is also seen. They are stable for smaller $a$ till $a=-\infty$. Thus we may conclude that orange curve can be taken as a almost gauge independent physical branch.

Roughly speaking the region $-0.5 \simlt a \simlt 2.8$ is a dangerous zone where multiple branches needed to cover the orange curve or good branch disappear at a certain point.\footnote{Movie files of the potential for varying $a$ are available at {\tt http://hep1.c.u-tokyo.ac.jp/\~{}kato/sft/}.} It is clear from the figure that the  Siegel gauge ($a=0$) is at the edge of the zone. The pathological behavior in this region seems to be coming from the influence of nearby gauge horizon. For example in free theory $a=1$ is a gauge non-fixed point. After switching on the interaction there still be a gauge invariance at $\phi=\psi_i=0$ configuration. Indeed we can see in the plot that three branches degenerate at the origin for $a=1$. For the non-zero values of fields a degeneration point would shift from $a=1$.

We can estimate the value of $a$ at which the gauge fixing fails in the interacting case. The level truncated potential $V^{(2,6)}$ has no exact gauge invariance even before the gauge condition is imposed so that we can solve the equation of motion derived from this action. Then we can find a vacuum solution~\cite{Rastelli:2000iu} near the vacuum of gauge-fixed case. In the level (2,6) truncation an approximate gauge transformation has only one parameter since there is only one state $b_{-2} \ket{\!\downarrow}$ in level 2 with ghost number 0. We can easily see that around $a=1.85$ the gauge slice determined by $a$ becomes parallel to the tangential direction of the gauge orbit piercing through the vacuum solution. Actually there are multiple local minima near the vacuum at $a=1.8$ in Fig.\ref{fig1}.

Note that in the plot for $a=-2$ a vertical branch crosses the physical branch near the origin. This, however, is not a degeneration since the values of fields other than $\phi$ are different from each other.

\begin{figure}[htbp]
\begin{center}
\includegraphics[scale=1.1]{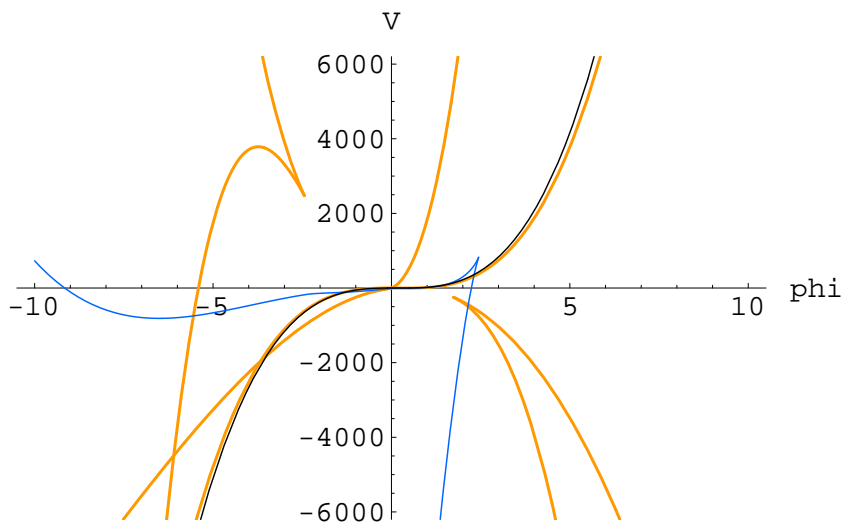}
\caption{{\bf Branches at $a=\infty$ (orange), $a=0$ (blue) and level 0 (black).}}
\label{fig2}
\end{center}
\end{figure}

Now at this point, one may wonder how wide is the region where the physical branch at $a=\infty$ behaves smoothly beyond the one shown in Fig.\ref{fig1}.
So next let us look at the global structure of branches. In Fig.\ref{fig2} we show all the branches at $a=\infty$ (orange curve). There is one physical branch which is all along the curve of level 0 (black curve), while the other branches are always independent and non-degenerate to the physical one.
There also shown is the curve of Siegel gauge (blue curve) which has several branch points on the curve and coincides with $a=\infty$ curve only in the small region around $g\bar{\kappa}\phi\sim1$.

\subsection{Higher level results}

\begin{figure}[htbp]
\begin{center}
\begin{picture}(460,200)(50,0)
\put(0,0){\includegraphics[scale=.9]{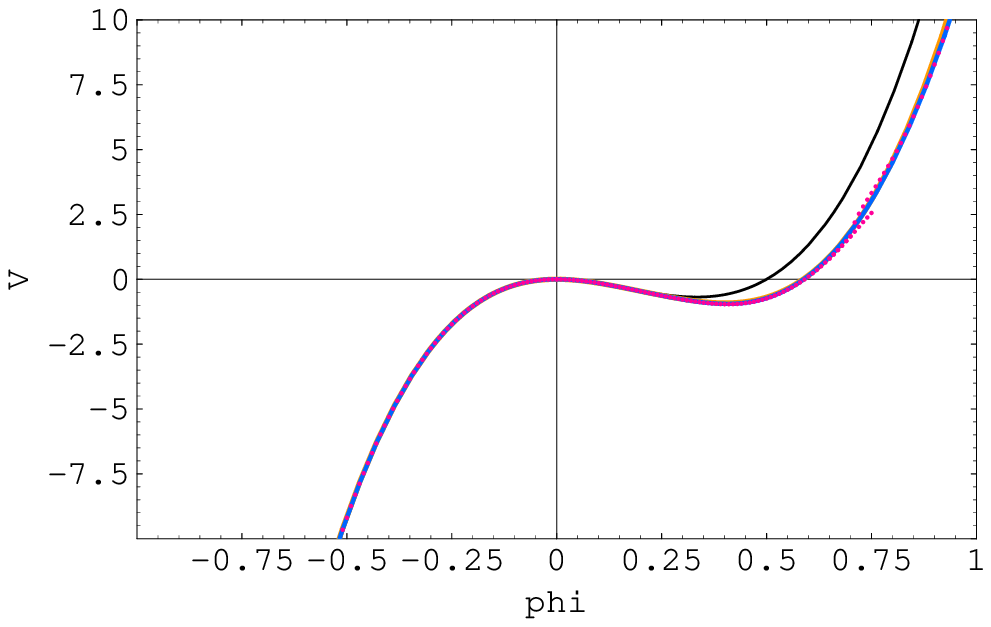}}
\put(270,0){\includegraphics[scale=.9]{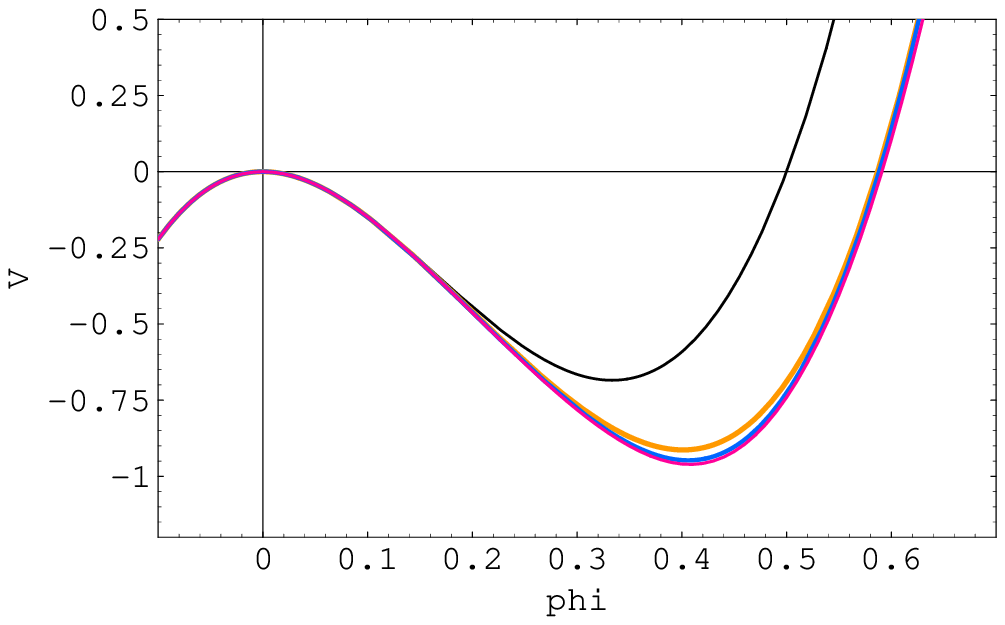}}
\end{picture}
\caption{{\bf Tachyon potential at $a=\infty$ in level 0 (black), 2 (orange), 4 (blue) and 6 (rose) truncations.}}
\label{fig3}
\end{center}
\end{figure}

At level 2, we have at most four real branches at fixed $\phi$ since we solve two quadratic equations with respect to two $\psi_i$'s. For higher levels we may have enormous number of branches in principle. So practically we only look for physical branch beyond level 2. We actually have calculated in level (4,12) and (6,18) truncations at $a=\infty$ whose results are summarized in Fig.\ref{fig3}.

We can see from the figure that the potential shape is quite stable. In particular, in negative $\phi$ region the level (2,6) approximation is already quite good and gets little correction from higher level. The higher level correction is only apparent near the local minimum of the potential.

Note that the branch structure is seen for (6,18) around $g\bar{\kappa}\phi\sim0.75$ where the branch coming from lower value of $\phi$ starts to deviate and another branch redeems it instead. This behavior is quite different from the Siegel gauge in which the branch stops at a certain point and no other branch shows up. In addition, the higher the level goes up the smaller the effective region of the branch becomes in the Siegel gauge.

\section{Gauge independence of the vacuum}

\begin{figure}[htbp]
\begin{center}
\includegraphics{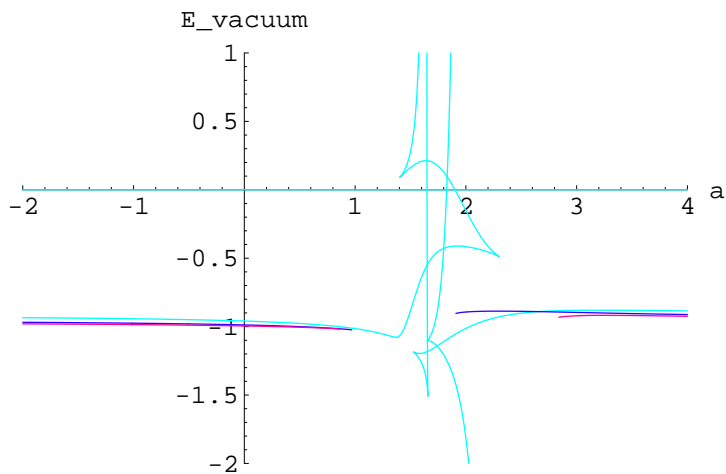}
\caption{{\bf Energy of the vacuum solution versus $a$ in level 2 (sky), 4 (blue) and 6 (rose) truncations.}}
\label{fig4}
\end{center}
\end{figure}

Now in this section let us concentrate on the local extremum of the potential, {\it i.e.}, the solutions of the equations of motion. 
In Fig.\ref{fig4}, we plotted the $a$-dependence of the normalized energy of the solutions. All the solutions in the plotted area are shown for level (2,6) truncation and only vacuum solution for (4,12) and (6,18). We checked the other solutions in level (2,6) outside the area are highly $a$-dependent, hence gauge-artifacts.

We see from the figure that away from the dangerous zone the vacuum solution is gauge independent quite stably, so that the choice of solution is justified.
The dangerous zone is slightly narrower than the potential analysis in the previous section. This is because only extrema are relevant here while the potential should be defined for any values. In this sense the Siegel gauge is not safe for almost all the value of $\phi$ but around the vacuum.

\begin{figure}[htbp]
\begin{center}
\includegraphics{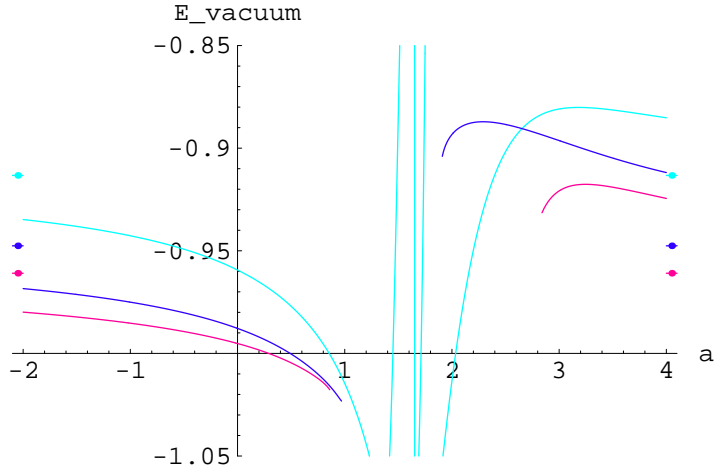}
\caption{{\bf Magnified view of Fig.\ref{fig4} around the energy $\sim -1$.} Points at leftmost and rightmost ends are the values at $a=\infty$.}
\label{fig5}
\end{center}
\end{figure}

In order to see how higher level corrections affect the vacuum energy, the part around the normalized energy $\sim-1$ is magnified in Fig.\ref{fig5}.
The energy value at $a=\infty$ in each level truncation is shown as a spot in both side of the plot.

At first sight the energy value of Siegel gauge ($a=0$) is lower than $a=\infty$. We should, however, be careful about that we obtain even lower value than the Siegel gauge in approaching the dangerous zone. In fact for $a=0.5$ where we have very similar behavior of the potential as the Siegel gauge, the energy value overshoots the desired value $-1$ (see Table~\ref{table1}), which is also known for the higher level analysis in Siegel gauge~\cite{Gaiotto:2002wy}. It will be interesting to see whether the same phenomenon happens or not in much higher level for $a=\infty$.

\begin{table}[htbp]
\begin{center}
\begin{tabular}{|r|c|c|c|c|c|c|c|c|}
\hline
&\multicolumn{8}{c|}{Level}
\\\cline{2-9}
& \multicolumn{2}{c|}{(0,0)} & \multicolumn{2}{c|}{(2,6)} & \multicolumn{2}{c|}{(4,12)} & \multicolumn{2}{c|}{(6,18)} 
\\
\cline{2-9}
\multicolumn{1}{|c|}{$a$} & $g \bar{\kappa} \phi_{\rm vac}$ & $E_{\rm vac}/T_{25}$ & $g\bar{\kappa}  \phi_{\rm vac}$ & $E_{\rm vac}/T_{25}$ &
$g\bar{\kappa} \phi_{\rm vac}$ & $E_{\rm vac}/T_{25}$ & $g \bar{\kappa} \phi_{\rm vac}$ &  $E_{\rm vac}/T_{25}$
\\
\hline\hline 
$\infty$ & &  &  0.40112 &$-0.91328$ 
& 0.40601  & $-0.94758$ & 0.40785 & $-0.96094$ 
\\
\cline{1-1}\cline{4-9}
$4.0$ & &  & 0.41188 & $-0.88520$ 
& 0.41753 & $-0.91189$ & 0.42401 & $-0.92449$
\\
\cline{1-1}\cline{4-9}
$0.5$ &  0.33333 &$-0.68462$  & 0.39843 & $-0.97704$ & 0.40076 & $-1.00030$
&  0.40012 & $-1.00459$
\\
\cline{1-1}\cline{4-9}
$0.0$ & & & 0.39766 &$-0.95938$ & 0.40072 & $-0.98782$ &
0.40038 & $-0.99518$ 
\\
\cline{1-1}\cline{4-9}
$-2.0$ & & &  0.39828 & $-0.93477$ 
& 0.40211 & $-0.96842$ & 0.40242 & $-0.97989$ 
\\
\hline
\end{tabular}
\caption{The field value and the energy of tachyon vacuum for various $a$.}
\label{table1}
\end{center}
\end{table}

\section{Discussions}

We have analyzed the tachyon condensation in the level truncation approach by the use of new covariant gauge. As we have seen in section 3, if we take the gauge parameter $a$ sufficiently away from the dangerous zone, the tachyon potential has a well-behaved physical branch which is all along the level 0 potential with an appreciable correction. In the same region the tachyon vacuum solution is almost gauge independent. The Siegel gauge ($a=0$) is turned out to be on the edge of dangerous zone so that we should be careful about its validity depending on the field value.

We suggest that the $a=\infty$ gauge is quite stable and reliable. As is shown quite generally in Ref.\cite{AK} the action is largely simplified in this gauge, so that not only numerical analysis but also analytic investigation may be promising. Indeed the number of terms in the action of scalars at $a=\infty$ is roughly 80 percent of that of Siegel gauge in each level.

Even if we include derivative terms in order to study space-time dependent solutions, still $a=\infty$ gauge has a simple structure, since $\omega_j$'s have no derivatives in their kinetic terms. Note that $a=\infty$ gauge condition does not touch $\omega_j$'s while the Siegel gauge removes all $\omega_j$'s. Hence it is clear that the former has great advantage over the latter in the space-time dependent analysis.

Final words should go to the approximate BRST invariance~\cite{Hata:2000bj} which measures the approximate gauge invariance at a chosen gauge slice. Apparently the $a$-independence of the solution is closely related to the BRST invariance even in the level truncation. So it might be interesting to compare these two numerically in varying $a$.

\section*{Acknowledgements}
We would like to express sincere thanks to T.~Takahashi for providing us his valuable data on the level truncation which was much helpful for us.
The work is supported in part by the Grants-in-Aid for Scientific Research (17740142~[M.A.], 13135205 and 16340067~[M.K.]) from the Ministry of Education, Culture, Sports, Science and Technology (MEXT) and from the Japan Society for the Promotion of Science (JSPS).


\end{document}